# SocialScope: Enabling Information Discovery on Social Content Sites

Sihem Amer-Yahia
Yahoo! Research
New York, NY
sihem@yahoo-inc.com

Laks V.S. Lakshmanan
Univ. of British Columbia
Vancouver, Canada
laks@cs.ubc.ca

Cong Yu
Yahoo! Research
New York, NY
congyu@yahoo-inc.com

## ABSTRACT

Recently, many content sites have started encouraging their users to engage in social activities such as adding buddies on Yahoo! Travel and sharing articles with their friends on New York Times. This has led to the emergence of *social content sites*, which is being facilitated by initiatives like OpenID[1] and OpenSocial[2]. These community standards enable the open access to users' social profiles and connections by individual content sites and are bringing content-oriented sites and social networking sites ever closer. The integration of content and social information raises new challenges for *information management and discovery* over such sites. We propose a logical architecture, named SocialScope, consisting of three layers, for tackling the challenges. The *content management* layer is responsible for integrating, maintaining and physically accessing the content and social data. The *information discovery* layer takes care of analyzing content to derive interesting new information, and interpreting and processing the user's information need to identify relevant information. Finally, the *information presentation* layer explores the discovered information and helps users better understand it in a principled way. We describe the challenges in each layer and propose solutions for some of those challenges. In particular, we propose a uniform algebraic framework, which can be leveraged to uniformly and flexibly specify many of the information discovery and analysis tasks and provide the foundation for the optimization of those tasks.

## 1. INTRODUCTION

Web 2.0 is leading to an increasing integration of content information with the social information (profiles, connections and activities) of users, giving rise to *social content sites*. Sites like Flickr and del.icio.us started out as such sites by enabling users to tag and share contents like photos and bookmarks. More recently, however, sites that started as pure content oriented or pure social networking focused are increasingly marching toward such an integration. For example, content sites like Amazon and Yahoo!Travel are becoming more social: users can now become friends, share content with each other, and tag the content with their own descriptions. Similarly, social sites like MySpace and Facebook are adding more content: users can add contents like photos and news items to their personal spaces, making the site more practically useful in their daily lives.

We envision this integration of social sites and content sites to be greatly helped by initiatives like OpenID and OpenSocial. They can now collaborate with each other to form *virtual* social content sites, where the social sites manage the user's social life and the content sites manage the detailed content information. We are already witnessing this in many domains. For example, on most online news sites (e.g., New York Times), each article is accompanied by buttons corresponding to Facebook, del.icio.us, etc., which allow you to quickly post the article to your favorite social site and share it with friends. Together, the social site(s) and the content site form a powerful virtual social content site that can engage a larger number of users much more deeply than each individual site.

Another important trend is the *increasing structural richness of information* about the users and the content. On one hand, social sites are knowing more about us through the rich information we voluntarily provide (e.g., name, interests, etc.). On the other hand, content sites are generating more structured information as a result of advances in information extraction and wiki-style mass collaborations. Only a few years ago, almost all the Wikipedia pages were pure text articles. Now, nearly all of the highly visited ones contain some structured information (e.g., Infoboxes) and are organized within a set of loosely defined category hierarchies.

The ways in which those sites help their users discover information, however, have evolved little from the traditional keyword-based search paradigm. The rapidly growing social graphs underlying those social content sites are rarely leveraged to better serve the user's information needs. There is still a dichotomy between information retrieval, which focuses on locating information that is semantically relevant to a user query, and information recommendation, which focuses on identifying information a user might prefer based on her social profile activities and those of her social connections. Finally, results are still ranked and presented in a predominantly list-based fashion, not taking advantage of the rich structure and social provenance embedded in them.

In this paper, we identify the research challenges and opportunities associated with managing and discovering information on social content sites. We present an architectural vision, SocialScope, as a platform where those challenges can be addressed. But first, we provide some motivating examples in the context of a real-world social content site, Yahoo!Travel.

---

[1] http://www.openid.net/
[2] http://www.opensocial.org/

|  | general (e.g., things to do) | categorical (e.g., family) | specific |
|---|---|---|---|
| with locations | 32.36% | 22.52% | 8.37% |
| w/o locations | 21.38% | 5.34% | |

**Table 1:** Summary Statistics of 10 Million Y!Travel Queries.

## 2. CASE STUDY WITH YAHOO! TRAVEL

Y!Travel[3] is a typical content site that is gradually evolving into a social content site. Initially built as a portal on travel destinations, it has been incorporating various social features including allowing users to tag travel destination and showing them similar travelers. It has also been interacting with Y!Local and Flickr to provide more structured information and social data about destinations. These days, users visit Y!Travel to look for information about travel destinations as well as to learn about their friends and other travelers. We briefly describe the data and queries in Y!Travel.

**Y!Travel Data**: Y!Travel maintains a comprehensive set of travel objects: cities, restaurants, etc., each with its own structure. Various semantic links are established between objects. For example, Fisherman's Wharf and San Francisco are connected through geographical containment. Users on Y!Travel provide detailed information about themselves, including self-tags, interests, etc., and they are also connected in various ways. For example, they can be friends on Flickr or contacts on the Instant Messenger network. Finally, users browse travel objects, tag them with keywords, and provide ratings and reviews on them, creating connections between users and objects.

**Y!Travel Queries**: Users interact with Y!Travel through a search interface, where they enter a set of keywords and obtain a list of travel objects considered relevant to their queries. We conducted a comprehensive analysis on 10 million recent Y!Travel queries to better understand the user behavior. The results are summarized in Table 1. By leveraging the domain knowledge we have about geographical locations and travel destinations, we detect location terms in queries and classify each query into three classes: *general, categorical, and specific*.[4] General queries are those containing terms like "things to do", "attraction", or just a location by itself. Over 50% of the queries fall into this class, and about 60% of those queries contain a location. Categorical queries refer to those containing terms like "hotel", "family", "historic", etc. About 30% of the queries fall into this class and a majority of them mention a location. Finally, there are also about 8% of the queries looking for specific destinations like "Disneyland" and "Yosemite Park".

The distributions of Y!Travel queries indicate that the main information needs of users are not specific destinations, but rather the set of *interesting* destinations among a large group that are loosely constrained by the general or categorical queries. This is in contrast with web search engines, where users are mostly searching for specific information. As a result, the search paradigm is a poor fit for Y!Travel because it is inherently hard to discriminate among a large group of results based purely on the query keywords. For example, almost all destinations in Y!Travel are *attractions*. To address this problem, Y!Travel manually creates guides of "things to do", "hotels", and "restaurants" for popular destinations, which are extensively browsed by users. However, it is impossible to manually create the "right" guide for each user on all destinations and queries. A new information discovery paradigm is therefore needed for Y!Travel and other similar social content sites. In the rest of this section, we describe our vision of this new paradigm through three hypothetical examples involving Y!Travel that are synthesized from our extensive conversations with actual users.

### 2.1 Motivating Examples

EXAMPLE 1. *John is in Denver for a conference. Having one day free, he visits* Y!Travel *and searches for "Denver attractions". John has in the past visited quite a few baseball fields on* Y!Travel *and has many friends on* Facebook *with interests in "baseball". With this knowledge,* Y!Travel *recommends to him "B's Ballpark Museum" (a small baseball museum in the suburb), "Coors Field" (home field of the Rockies), as well as the upcoming baseball game "Yankees vs Rockies" to be played at Coors Field, which is fetched from* Y!Sports.

Example 1 represents one out of three queries on Y!Travel. However, the traditional information retrieval approach fails for it because there are often many objects that are semantically relevant to John's query and no ranking mechanism (e.g., tf-idf measure) based on pure *semantic relevance* can differentiate them. It is therefore imperative for the system to incorporate *social relevance*, which considers John's social profile and connections, to decide which attractions he will prefer. Essentially, information discovery on social content sites requires the integration of two major paradigms: semantic relevance with respect to a query and social relevance in the spirit of recommendations. The former scopes the discovery to information relevant to John's current needs as expressed by him, while the latter identifies the information most appealing to John as a user. Indeed, "B's Ballpark Museum" may not be a major attraction, yet John, being a baseball fan, is likely to enjoy a visit to it. Example 1 further illustrates another important desideratum: the need to *retrieve relevant information from external social or content sites* that are physically and administratively separate from Y!Travel, which is becoming possible because of various initiatives like OpenSocial.

EXAMPLE 2. *Selma, a young musician with two babies, is planning a family trip to Barcelona. She searches for "Barcelona family trip with babies" on* Y!Travel. *As in John's case,* Y!Travel *searches for attractions that are semantically and socially relevant to Selma. While Selma is well-connected to her musician friends, very few of them have kids and are suitable for trip recommendation to Selma in this case. Instead,* Y!Travel *analyzes her other friends who have made similar family trips before and uses them as the social basis for recommending baby-friendly attractions like the "Parc de la Ciutadella".*

Selma's example illustrates the importance of *analyzing the social connections of users* and choosing the right subset of the connections as the basis for discovering socially-relevant results. Unlike in John's case, Y!Travel has to understand the distinct groups of friends that Selma has and pick the right group for her family oriented query. This analysis, however, is non-trivial since the nature of social activities and connections are often implicit and noisy. Determining whether a social connection is suitable for answering a particular query is a significant challenge for a social content site and may even require an interaction with the user. Furthermore, it is not always possible or necessary to "personalize" social relevances. Even if Selma does not have any friend with young babies, Y!Travel should still be able identify a group of "experts" on the topic to help answer Selma's query. This would require extensive data analysis to identify topics within the data and users with expertise on the topics.

EXAMPLE 3. *Alexia is a high school student planning a summer field trip for an assignment from her history class. She comes to* Y!Travel *and searches for "American history" to find places for research on the subject. As in previous examples,* Y!Travel

---

[3] http://travel.yahoo.com/

[4] There are about 10% of the queries that we were unable to classify.

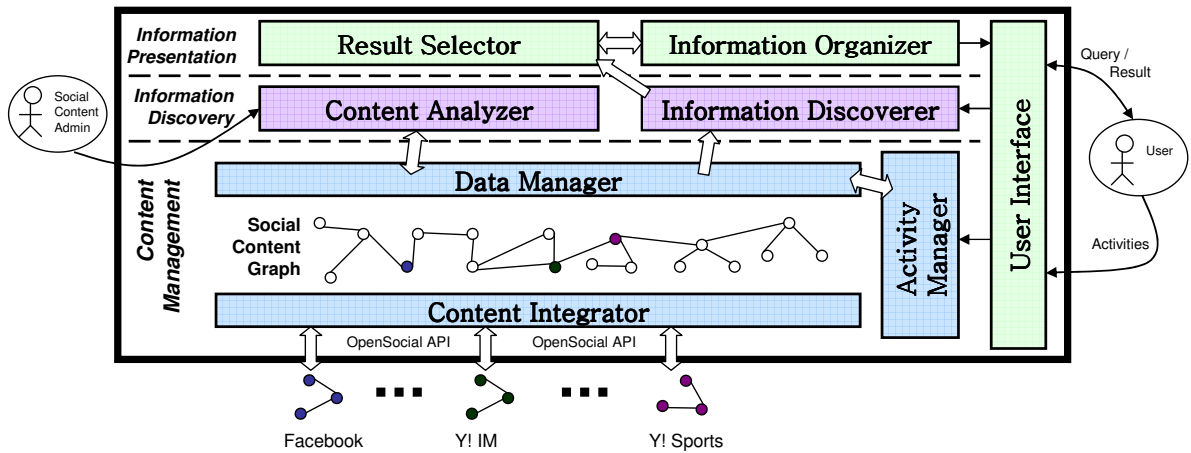

**Figure 1:** Architecture of SocialScope.

*leverages Alexia's social information and finds a set of semantically and socially relevant places. However, the results this time contain places from throughout the country and fall into many different topics. Recognizing the chaotic nature of the results,* `Y!Travel` *moves away from the traditional list-based presentation, and automatically groups results along multiple dimensions: geographical or organized based on who—her classmates in the history class or her friends on the soccer team—endorse it. Furthermore,* `Y!Travel` *analyzes the result destinations and present related topics (e.g., Independence War) and users (e.g., Jane, who left comments on many result destinations) to Alexia.*

Example 3 illustrates another important aspect in information discovery on social content sites: *result presentation*. Similar to Alexia's case, there can be many equally relevant (semantically and socially) results to a user and her query, and alternative presentation mechanisms need to be employed for the users to effectively explore the results. For example, grouping can be accomplished based on the rich structure information associated with each object in a way similar to faceted search. More interestingly, in social content sites, each result also has an associated social provenance that can be explored for more sophisticated grouping. Furthermore, users like Alexia are often not just specifically looking for objects: they are also interested in exploring other information (e.g., similar users and associated topics) related to their information need. `Y!Travel` needs to detect when such explorations are warranted and how to facilitate them.

### 2.2 Insights from the Examples

As all three examples have emphasized, effective social content discovery calls for an *integration of three major paradigms: keyword search, database-style querying, and recommendations.* The search/query paradigm helps users express their need and narrow down the discovery scope, while the recommendation paradigm enables the system to guide and expand the discovery process socially. The fact that social and semantic relevances play an equally important role distinguishes this from *personalized search* [21], where personalization is achieved by re-ranking semantically relevant results in a post-processing step.

Second, *the social information on social content sites are much more complex* than in traditional recommendation systems [24]. Instead of being characterized simply by what items they have read or bought previously, users can participate in various social activities and establish connections with different semantics. These connections need to be analyzed and selectively applied to help the discovery process.

Finally, the fact that most user queries are exploratory in nature calls for *effective ways to present the results* to facilitate information exploration. While faceted search [14] has made strides in helping users explore query results, it does not account for social provenances. Discovering the most effective groupings, either based on structural attributes or on social relevance, is a significant new challenge.

### 3. ARCHITECTURAL VISION

Figure 1 describes the logical architecture of `SocialScope`. At its core is the *social content graph*, which represents users, objects, and various connections among them. Information in the graph may be *locally owned* (e.g., destinations in `Y!Travel`), *externally integrated* (e.g., friendship connection obtained from `Facebook`) or *derived* (e.g., links describing similarities between users). Three layers, *Content Management*, *Information Discovery*, and *Information Presentation*, form the entire `SocialScope` system, and we briefly describe each next.

**Information Discovery** (Section 5): This layer consists of two components: *Content Analyzer* and *Information Discoverer*. The Content Analyzer derives new nodes (e.g., topics) and links (e.g., similarities between users) through various analyses (e.g., Latent Dirichlet Allocation [8], association rule mining [3]) of the raw social content graph in an off-line fashion. Those analyses can be specified and triggered automatically by the system itself or by a *Social Content Administrator*. The Information Discoverer parses the user query, constructs its internal representations (based on various semantic and social relevance computations), and evaluates them on the social content graph. The result is a social content subgraph, called *Meaningful Social Graph* (MSG), that is semantically and socially relevant to a given user and query. One major vision of `SocialScope` is to enable *uniform manipulation of social content graphs*, leading to declarative, flexible, and optimizable graph analysis and information discovery processes. We accomplish this by proposing a logical algebraic framework for social content graph manipulation.

**Content Management** (Section 6): This layer handles two main tasks. First, it facilitates the incorporation of social information from remote sites through *Content Integrator*. This has become increasingly important as open standards like OpenSocial become widely accepted, which allow the core social content sites to leverage the large amount of information within social sites. The second task is the maintenance and retrieval of the social content graph through the *Data Manager*, which abstracts away the physical im-plementation of the graph. In addition, given that much social content is created and maintained externally, Data Manager needs to make decisions on when and how to refresh parts of the social graph efficiently. The *Activity Manager* helps in that regard by categorizing users based on their activities.

**Information Presentation** (Section 7): This layer provides a comprehensive result exploration framework. It admits as input the MSG from the Information Discovery layer and dynamically organizes the results for effective exploration by the user. There are two key primitives: grouping and ranking, managed by Information Organizer and Result Selector, respectively. The former identifies appropriate (structural or social) criteria for grouping results, while the latter identifies appropriate mechanisms for ranking and selecting results within or across groups. When multiple presentation groups are available, Information Organizer also makes decisions on which group is more relevant to the user and her current information needs.

Let's begin by introducing the data and query models adopted by SocialScope.

## 4. DATA AND QUERY MODEL

**Nodes and Links**: We adopt a graph model for representing social content. Intuitively, nodes in the graph represent *physical and abstract entities* like users and topics, and links represent *connections and activities* between entities such as friendship and tagging actions. Each node or link has a unique id. It is worth noting that the graph model described here is a logical model that is not tied to any specific physical implementation.

Nodes and links contain structural attributes, including a mandatory type attribute. We adopt a flexible (i.e., schema-less) typing system and allow the type attribute to have multiple values. For example: $n_1$ = {id=1; type='user, traveler'; name='John'} and $n_2$ = {id=2; type='item, city'; name='Denver'; keywords='skiing'} are two nodes representing our traveler John and the city Denver, respectively, in Example 1. Similarly, $l_{12}(n_1, n_2)$ = {id=12; type='act, tag'; date='2008-8-2'; tags='rockies baseball'} is a link recording the activity that John tagged Denver with tags 'rockies baseball'. Our typing system gives us the flexibility of creating new types through content analysis. We also maintain an evolving catalog of basic types, including user, item, topic, group for nodes and connect (e.g., friend), act (e.g., tag, review, click, etc.), match, belong for links. Those basic types are adequate for modeling most of the social content sites we have encountered.

**Social Content Graphs**: We model an instance of a social content site as a *social content graph*. A social content graph consists of nodes and links as described above. It is sometimes convenient to view the social content graph as an overlay of sub-graphs, namely the *activity graph*, which maintains users' activities on items, the *network graph*, which maintains social connections, and the *topical graph*, which maintains links from users or items to derived semantic groups or topics.

**Queries**: Users interact with SocialScope by specifying a (possibly empty) query on content and structure[5]. Structural predicates are interpreted in the usual Boolean sense, while content conditions are used to compute semantic relevance which, combined with social relevance, results in a single relevance score. The system generates recommendations within the scope defined by the query, treating the structural predicates as the constraints defining the scope. When the structural predicates are absent in the query, only semantic relevance and social relevance are taken into account. And when a query is empty, only social relevance is accounted for.

## 5. INFORMATION DISCOVERY: AN ALGEBRAIC FRAMEWORK

Developing *an efficient and flexible mechanism to manipulate the social content graph* is a major goal of the Information Discovery layer. While social network graphs have been the subject of a number of social network analyses [23, 27, 29] and social activity graphs have been leveraged in various recommendation algorithms [2, 24], most of those works are designed for simple graphs (i.e., no complex structures on nodes or links) and adopt ad-hoc methods. This leaves the system with few opportunities for reuse, customization and optimization. Furthermore, while searching for objects based on content relevance has been extensively studied before, it has never been integrated into a social context in a principled way. We believe a uniform algebraic framework is needed to manipulate the kind of complex social content graphs we encounter in social content sites and to provide flexibility in the manner in which information is analyzed and discovered.

We thus propose a logical algebra that is capable of expressing sophisticated tasks for data analysis and discovering socially and semantically relevant results. Each operator in the algebra takes social content graphs as input and outputs a social content graph. In the next section, we present our algebra formally and demonstrate the expressive power of the algebra by showing a comprehensive set of tasks that can be expressed in the algebra.

### 5.1 Unary Operators

At the core of the algebra are two unary selection operators: Node Selection ($\sigma^N_{\langle C, S \rangle}$) and Link Selection ($\sigma^L_{\langle C, S \rangle}$). Both operators take a condition $C$ and an optional scoring function $S$ as parameters, and a (social content) graph as input. The condition $C$ consists of a list of structural conditions (e.g., {type='city', rating $\geq$'0.5'}) and a set of keywords (e.g., 'Denver attraction'). Satisfaction of the structural conditions by a node is defined in the obvious manner: a node $v$ is said to satisfy a structural condition of the form att=$val_1, ..., val_k$, if the set of $v$'s values for att is a superset of the values $\{val_1, ..., val_k\}$. When an optional scoring function $S$ is specified as an input parameter, a score is generated using $S$ for each node based on how well its content matches the keywords in $C$. If no scoring function is specified, but $C$ includes keywords, a default scoring function is used for generating the score. Finally, Node Selection outputs a null graph consisting of nodes (and no links) of the input graph that satisfy the node condition $C$. And a score is generated (by $S$) and attached to each node in the output graph. More formally:

DEFINITION 1 (NODE SELECTION). $\sigma^N_{\langle C, S \rangle}(G) =$ $\{v, v.score = S(v) \mid v \in nodes(G) \land v \text{ satisfies } C\}$. □

Link Selection is defined in an analogous manner, with the same format specification and satisfaction definition for condition $C$. Link Selection outputs a subgraph of the input graph induced by those links satisfying the selection condition $C$. And a score is generated by the optional scoring function $S$ and attached to each link within the output graph. More formally:

DEFINITION 2 (LINK SELECTION). $\sigma^L_{\langle C, S \rangle}(G) =$ $\{\ell, \ell.score = S(\ell) \mid \ell \in links(G) \land \ell \text{ satisfies } C\}$. □

Note that in the examples that will follow, we often omit the scoring function for clarity.

### 5.2 Basic Binary Operators

Also in the core framework are the binary set operators. We define the three common operators, Intersection ($\cap$), Union ($\cup$), Minus ($\setminus$), as follows:

---

[5]Here and elsewhere, the term structure refers to the attribute/value pairs associated with nodes and links.

DEFINITION 3 (SET-THEORETIC OPERATORS). Let $G_1$ and $G_2$ be two social content graphs originated from the same social content site. $G_1 \cup G_2$, $G_1 \cap G_2$, and $G_1 \setminus G_2$ are defined as: $nodes(G_1 \oplus G_2) = nodes(G_1) \oplus nodes(G_2)$ and $links(G_1 \oplus G_2) = links(G_1) \oplus links(G_2)$, where $\oplus$ is one of $\cup, \cap, \setminus$, and nodes and links with the same id are consolidated in the output graph. □

As an example, a link belongs to $G_1 \setminus G_2$ if and only if it is in $G_1$ but not in $G_2$. Note that a link in $G_1 \setminus G_2$ must necessarily be incident on nodes which appear in $G_1$ but not in $G_2$.

**Remarks**: First, in all the definitions above, nodes and links are matched on the basis of their id, as a result, graph isomorphism is not an issue. Second, we note that the operator $\setminus$ can be defined in more than one way. According to the definition above, $G_1 \setminus G_2$ is the subgraph of $G_1$ induced by those nodes of $G_1$ which are not present in $G_2$. Thus, all links in $G_1 \setminus G_2$ are necessarily those for which both endpoints are present in $G_1$ but not in $G_2$. We call this the *Node-Driven Minus* operator. Below, we give an alternative, link-driven, definition of the Minus operator, denoted '$\setminus\cdot$'.

DEFINITION 4 (LINK-DRIVEN MINUS). Let $G_1$ and $G_2$ be two social content graphs originated from the same social content site. $G_1 \setminus\cdot G_2$ is defined as: $links(G_1 \setminus\cdot G_2) = links(G_1) \setminus links(G_2)$; $nodes(G_1 \setminus\cdot G_2)$ consists precisely of those nodes which are induced by the set of links in $links(G_1 \setminus\cdot G_2)$. □

As an example of the difference between Node-Driven and Link-Driven Minus operators, consider $G_1 = \{(a,b),(a,c),(b,c)\}$ and $G_2 = \{(a,b)\}$. $G_1 \setminus G_2$ is a null graph containing only node $c$ and no links. On the other hand, $G_1 \setminus\cdot G_2$ contains all the three nodes $a, b, c$ and the links $(a, c)$ and $(b, c)$. We note in Lemma 1 that the Link-Driven Minus operator can be expressed with a combination of Node-Driven Minus operator and Semi-Join operator (to be described later).

LEMMA 1. Operator $\setminus\cdot$ can be expressed using operators $\setminus$ and $\ltimes$. (Proof omitted for clarity.) □

## 5.3 Advanced Binary Operators

Next, we introduce more sophisticated binary operators. Operator Composition $G_1 \odot_{\langle \delta, \mathcal{F} \rangle} G_2$ takes a directional condition $\delta$ and a composition function $\mathcal{F}$ as parameters and produces a graph induced by new links that are composed from links in $G_1$ and $G_2$. Input links to be composed are selected if they satisfy the directional condition $\delta$. And each new link in the output is attached with attributes generated by the function $\mathcal{F}$. The directional condition $\delta$ consists of two link directions, $d_1$=src|tgt and $d_2$=src|tgt, corresponding to links in $G_1$ and $G_2$, respectively. E.g., $\delta$=(src, tgt) means two links are composed if and only if the source node of the $G_1$ link matches the target node of the $G_2$ link, where two nodes match if and only if they have the same id.

**Composition Function**: Intuitively, the composition function $\mathcal{F}$ combines the attributes of input links and generates new attributes for the output link produced by composition. Since there can be a variety of user-defined composition functions, we focus on the their core requirements here. First, a composition function must accept as input two groups of attributes (and their values) corresponding to the two input links. These attributes may be link attributes or node attributes. Second, a composition function must produce as output a group of uniquely named attributes (and their values) to be associated with the output link. If a function satisfies both requirements, we consider it in the class of **CF**. Formally, the composition operator is defined as follows (note that $\delta_{\bar{d}_i}$ indicates the opposite direction of $\delta_{d_i}$):

DEFINITION 5 (COMPOSITION). Let $G_1$ and $G_2$ be two social content graphs originated from the same social content site. $G = G_1 \odot_{\langle \delta, \mathcal{F} \rangle} G_2$, where $\mathcal{F}$ is a function in the class **CF**, is defined as:

- $\forall u, v, \ell[u, v \in nodes(G), \ell \in links(G)$ if and only if $\exists \ell_1 \in links(G_1), \ell_2 \in links(G_2)$ s.t. $u = \ell_1.\delta_{\bar{d}_1} \wedge v = \ell_2.\delta_{\bar{d}_2} \wedge \ell_1.\delta_{d_1} = \ell_2.\delta_{d_2} \wedge \ell.src = u \wedge \ell.tgt = v]$.

- $\ell.\{\text{att}_1, \text{att}_2, ...\} = \mathcal{F}(\ell_1, \ell_2)$ □

In contrast with composition, operator Semi-Join $G_1 \ltimes_\delta G_2$, produces a subgraph of $G_1$ induced by the $G_1$ links that match the links in $G_2$. Again, links to be joined are selected if they satisfy the directional condition $\delta$. Both Composition and Semi-Join "connect" links in their input graphs. However, composition generates new links while semi-join simply filters away unwanted links. As a special case, when $G_1$ ($G_2$) is a null graph (i.e., no links), we set $d_1$ (resp., $d_2$) to src. Formally, we have:

DEFINITION 6 (SEMI-JOIN). Let $G_1$ and $G_2$ be two social content graphs originated from the same social content site. $G = G_1 \ltimes_\delta G_2$ is defined as:

- $\forall \ell[\ell_{src}, \ell_{tgt} \in nodes(G), \ell \in links(G)$ if and only if $\ell \in links(G_1) \wedge \exists \ell_2 \in links(G_2)$ s.t. $\ell.\delta_{d_1} = \ell_2.\delta_{d_2}]$. □

EXAMPLE 4 (SEARCH). We illustrate how the search task, *"Find John's friends who have visited travel destinations near Denver and all their activities"*, can be accomplished. Given the social content graph $G$ and John's node id 101, we proceed as follows: John's network is $G_1 = \sigma^L_{C_2}(G \ltimes_{(src, src)} \sigma^N_{C_1}(G))$, where $C_1$ is id=101 and $C_2$ is type='friend'. Users who visited places near Denver are captured by: $G_2 = \sigma^L_{C_4}(G \ltimes_{(tgt, src)} \sigma^N_{C_3}(G))$, where $C_3$ is {type='destination', 'near Denver'} and $C_4$ is type='visit'. John's subset of friends who have visited places near Denver is then $G_3 = G_1 \ltimes_{(tgt, src)} G_2$, while places near Denver that are visited by John's friends are $G_4 = G_2 \ltimes_{(src, tgt)} G_1$. The union $G_5 = G_3 \cup G_4$ puts these two together. Activities by these friends are $G_6 = \sigma^L_{C_5}(G \ltimes_{(src, tgt)} G_3)$, where $C_5$ is type = 'act'. Finally, the union $G_7 = G_5 \cup G_6$ puts together John, his friends who have visited places near Denver, the places, and the friends' activities. □

## 5.4 Aggregation Operators

Aggregation is critical for most analysis tasks on social content graphs including model-based methods like Latent Dirichlet Allocation (LDA) [8]. One important feature of aggregation is the creation of new information (aggregation results) that need to be stored and maintained. Because of the rich structures of nodes and links, we can naturally incorporate aggregation results as new attributes. We define two operators Node Aggregation ($\gamma^N_{\langle C, d, att, \mathcal{A} \rangle}(G)$) and Link Aggregation ($\gamma^L_{\langle C, att, \mathcal{A} \rangle}(G)$), where $C$ is a link condition as described in Section 5.1, $d$=src|tgt is a directional constraint, att is the destination attribute whose value will contain the aggregation result, and $\mathcal{A}$ is the aggregation function.

**Aggregation Function**: Intuitively, the aggregation function takes as input a collection of links (and their associated attributes and values) and produces as output a value to be associated with the attribute att. We focus on two classes of aggregations: (i) the class **SAF** of set aggregation functions that map a set of links to a set of scalars, and (ii) the class **NAF** of numerical aggregation functions that map a set of links to a numerical scalar value. We formally define both classes below. Note that in the following definition, we use $x to denote a variable. When att is a set-valued attribute of a link $\ell$, the expression $\ell.\text{att} = \$x$ binds $\$x$ to one value of $\ell.\text{att}$ at a time.

DEFINITION 7 (SET AGGREGATE FUNCTIONS). Let $L$ be a set of links. An aggregation function $\mathcal{A}$ is in class **SAF** if and only if it is of the form $\{\$x \mid \ell \in L \& \ell.\text{att} = \$x\}$, which extracts values of the attribute att from every link in the input set $L$ and forms an output set of scalar values. □

As an example, let $L$ correspond to the set of links corresponding to a user's tagging actions. Let tags be the attribute of these links that contains the tags assigned to each item. The function $\{\$x \mid \exists \ell \in L : \ell.\text{tags} = \$x\}$ forms the set of all distinct tags assigned by the user to the items she has tagged.

DEFINITION 8 (NUMERICAL AGGREGATE FUNCTIONS). The class **NAF** of aggregation functions is defined as follows:

- Every arithmetic operation $+, -, \times, \div$ is in **NAF**;

- The constant functions **0** and **1** which map every scalar input to the constant 0 and 1 respectively are in **NAF**;

- Summation over a collection, i.e., $\sum_{x \in X} f(x)$, where $X$ is a collection and $f$ is a function in **NAF**, is in turn in **NAF**;

- Product over a collection, i.e., $\Pi_{x \in X} f(x)$, where $X$ is a collection and $f$ is a function in **NAF**, is in turn in **NAF**;

- **NAF** is closed under composition. □

It is easy to see that popular aggregate functions like summation, count, and average can be readily expressed in **NAF**. For example, to count the number of elements in a set $X$, we do the following: $COUNT(X) ::= \sum_{x \in X} \mathbf{1}(x)$. Other aggregate functions like minimum and maximum can also be expressed, although the details of the construction is omitted here for the clarity of presentation. Henceforth, we refer to the union of the classes **SAF** ∪ **NAF** as simply **AF**.

DEFINITION 9 (NODE AGGREGATION). Let $G$ be a social content graph, $\gamma^N_{\langle C,d,\text{att},\mathcal{A}\rangle}(G)$ produces a social content graph $G'$ that is isomorphic to $G$ and $\forall v \in G'$ if $\exists \ell \in G \land \ell$ satisfies $C \land \ell.d = v$, then $v.\text{att}=\mathcal{A}(\{\ell_i \in \text{links}(G) \mid \ell_i \text{ satisfies } C \& \ell_i.d = v\})$. □

Notice that the directionality parameter $d$ acts as a group-by attribute, in that all outgoing links from a node (or all incoming links to a node) are grouped together and aggregated. As an example of node aggregation, suppose $\mathcal{A}(L)$ simply counts the number of links in $L$. Let the condition $C$ be type='friend'. The expression $\gamma^N_{\langle C, 'src', \text{fnd\_cnt}, \mathcal{A}\rangle}(G)$ produces a graph that is isomorphic to $G$ except for every node that has one or more outgoing 'friend' links. For those nodes, an attribute fnd_cnt is generated to store the aggregate value, namely the number of friends, as computed by the aggregation function. Similarly, node aggregation can be used to assign an attribute tags_used to every user node, whose values include all the tags that have been used by the user.

The definition of the Link Aggregation operator is analogous to Node Aggregation, except for two major differences. First, link aggregation changes the structure of an input graph: it replaces a set of links between a given *src* and *tgt* node by a new link. Secondly, the result of the aggregate computation is assigned as a destination attribute of the newly created link.

DEFINITION 10 (LINK AGGREGATION). Let $G$ be a social content graph, $\gamma^L_{\langle C,\text{att},\mathcal{A}\rangle}(G)$ produces a social content graph $G'$ as follows:

1. Partition $\{\ell \mid \ell \in \text{links}(G) \land \ell \text{ satisfies } C\}$ on $\ell.src$ and $\ell.tgt$;

2. For each set of links $\mathcal{L}_{s,t}$ sharing the same source node $s$ and the same target node $t$, replace $\mathcal{L}_{s,t}$ with a new link $\ell_{s,t}$;

3. Attach an attribute att with $\ell_{s,t}$, with its value computed as $\mathcal{A}(\mathcal{L}_{s,t})$.

□

As an example of link aggregation, let $G_1$ be a graph containing users and their friends, and let $G_2$ be a graph containing users and cities that they have visited. Both these subgraphs can be extracted easily from an input social content graph corresponding to a social content site, in a manner similar to that illustrated in Example 4. Furthermore, let $G_3$ be the result of composing $G_1$ and $G_2$, where the composed links contain attribute type='user_friend_item' and are results of composing friend links in $G_1$ and visit links in $G_2$. The link aggregation $\gamma^L_{\langle C,\text{vst\_cnt},COUNT\rangle}(G_3)$, where $C$ is the condition type='user_friend_item', replaces each set of links sharing the same user node and the same city node by one new link. It then assigns an attribute vst_cnt to the new links, whose value is computed by counting the number of user_friend_item links from the user node to the city node in $G_3$.

Next, we describe a comprehensive example that represents the collaborative filtering strategy of recommendation.

EXAMPLE 5 (COLLABORATIVE FILTERING). We show how *collaborative filtering* can be expressed for recommending travel destinations to John. Given the social content graph $G$ and John's node id=101, we proceed as follows:

1. $G_1 \longleftarrow \sigma^L_{\text{type}='visit'}(G \bowtie_{src,src} \sigma^N_{\text{id}=101}(G))$. $G_1$ now contains user John and the places he has visited.

2. $G'_1 \longleftarrow \gamma^N_{\text{type}='visit',src,\text{vst},\mathcal{A}}(G_1)$, where $\mathcal{A}$ is a set aggregation function that collects the set of destinations that John has visited and stores that as attribute vst of node John.

3. $G_2 \longleftarrow \sigma^L_{\text{type}='visit'}(G \bowtie_{src,src} \sigma^N_{\text{id}\neq 101}(G))$, finding users other than John and the places they have visited.

4. $G'_2 \longleftarrow \gamma^N_{\text{type}='visit',\text{vst},src,\mathcal{A}}(G_2)$, collecting the set of destinations that each user (other than John) has visited and stores that as attribute vst of the user node.

5. $G_3 \longleftarrow G_1 \odot_{\langle\delta,\mathcal{F}\rangle} G_2$, where $\delta = (tgt, tgt)$ and $\mathcal{F}$ is a composition function that computes the Jaccard similarity between user John and every other user and assigns the result to the attribute sim on the links produced by composition. The attribute vst of each user contains the necessary information for $\mathcal{F}$ to compute the Jaccard similarity between John and other users. Notice that this step produces one link from John to another user for every common place visited by both. The value of sim on all these links is the same.

6. $G_4 \longleftarrow \gamma^L_{\text{sim}>0.5,\text{type},\mathcal{A}'}(G_3)$, where $\mathcal{A}'$ is an aggregation function that assigns the constant string value 'match' to the destination attribute type and retains the value of sim from any of the input links.[6] This step replaces each set of links from John to another user similar to him with a weight over 0.5 by a new link with type='match'.

---

[6]Notice that this is well defined.

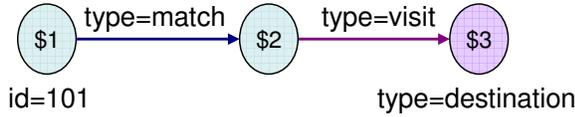

**Figure 2: Example of graph pattern for collaborative filtering.**

7. $G_5 \longleftarrow \sigma^L_{\text{type}=\text{'visit'}}(G \ltimes_{tgt,src} \sigma^N_{\text{type}=\text{'destination'}}(G))$. This step computes users and the destinations they have visited.

8. $G_6 \longleftarrow (G_4 \ltimes_{tgt,src} G_5) \odot_{\langle (tgt,src), \text{sim\_sc}, \mathcal{F}' \rangle} (G_5 \ltimes_{src,tgt} G_4)$. This step composes the two graphs: John and his similarity network friends (with similarity over 0.5), and users and the destinations they have visited. For each of John's similarity network friends, who has visited a destination, a new link is added between John and that destination. The function $\mathcal{F}'$ simply copies the value of attribute sim of the link from John to the user, on to the new link from John to the destination node and assigns this value to the attribute sim_sc.

9. $G_7 \longleftarrow \gamma^L_{C,\text{score},AVERAGE}(G_6)$. For each destination node, we replace the set of links from John to the destination node by one new link with an attribute score. The value of score is computed by taking the average of the sim_sc values on the links being aggregated.

Finally, destination nodes so obtained can be recommended to John on the basis of the computed score value. □

Often, aggregations can involve multiple links. For example, counting the number of each user's friends who have tagged at least five URLs with the term 'baseball', involves aggregation on friend and tagging links. This leaves us with two alternatives: allowing complex aggregation conditions like a *graph pattern* and therefore using fewer aggregation steps, or using more aggregation steps and therefore reducing the complexity of aggregation conditions.

As an example, we illustrate the use of graph patterns for expressing aggregations more concisely. In the above example, we used link aggregation confined to aggregating over links between a pair of nodes. As a result, we first had to create links from John to each destination node, one link for each similarity network friend of John that has visited that destination (Step 6). Then we had to perform a separate link aggregation to compute the score of each destination being recommended to John, as the average sim_sc value of the recommending user. Graph patterns make it possible to achieve these steps more concisely. Figure 2 depicts a graph pattern showing a 'match' link followed by a 'visit' link. First, we compute the union $G_4 \cup G_5$ of the graphs $G_4, G_5$ in Example 5, which contains John, his similarity network and the destinations they have visited. The operator $\gamma^L_{GP,\text{score},\mathcal{A}}(G_4 \cup G_5)$, where $GP$ is the graph pattern in Figure 2, creates a new link between John and a destination node whenever the latter is reachable from John by a match-visit link path. Only one link is created from John to the destination node, and the link is assigned an attribute score, whose value is computed as the average value of sim_sc on the match link of the set of match-visit paths from John to the destination node.

One of the research challenges we are pursuing is to *study the difference between the two approaches and identify the conditions under which one of the two approaches will be more effective.*

## 6. CONTENT MANAGEMENT

At the core of most social content sites, there are three major categories of data: *site content*, users' *social profiles and connections*, and users' *site-specific social activities*. Intuitively, site content is the content that users are interested in when they visit the social content site. Examples of such content include travel destinations in Y!Travel or URLs in del.icio.us. Social profiles and connections are the information regarding the users themselves (e.g., name, education, etc.) and their explicit social connections (e.g., friends, classmates, colleagues, etc.). Finally, site-specific social activities are the activities users perform on the site content. For example, in Y!Travel, users visit and browse destinations, while in del.icio.us, users bookmark URLs with tags.

How to effectively and efficiently manage the three categories of data is at the heart of challenges to be addressed by the Content Management layer of our SocialScope system. As a first step toward this goal, we describe and analyze three alternative management models for social content sites in Section 6.1. In Section 6.2, we discuss a detailed study on how the storage of large volumes of data can be optimized.

### 6.1 Models for Social Content Management

Logically, the *social content graph* is a single comprehensive graph that encompasses both content and social information relevant for the site. Physically, however, there are multiple models through which we can implement the social content graph, depending on how we maintain the social information.

**Decentralized Model**: In this model, each social content site maintains their own social information, including storing the user profiles and social connections, and effectively manages the entire social content graph internally. This is perhaps the most dominant model in the early days of Web 2.0, when sites like del.icio.us and Flickr were just starting, and they were all soliciting users' profiles and social connections on their own. This led to a set of decentralized social graphs, each residing in a different social content site, and collectively forming the global social graph.

This decentralized model provides social content sites with some obvious benefits, including full control over the entire data, which enables the site to perform comprehensive analysis on the social content graph, and increased exit cost on users, because they will have to leave their social connections behind and re-establish (probably the same) connections elsewhere if they decide to switch to a different site. It, however, has a couple of major problems. First, establishing a social graph with critical mass is incredibly difficult. Many social content sites can only provide strong user experience when they are able to leverage a large underlying social graph. For example, an event planning site is of no practical value if few of your family and friends are using it. This presents the *cold start* problem for many content sites that few of them can overcome. Second, social graph decentralization means it is necessary for users to establish their social connections multiple times on many different sites, even though most of those connections are the same. This creates unnecessary burdens on the users and deters them from adopting emerging social content sites.

**Closed Cartel Model**: With the emergence of several dominant social networking sites, the Closed Cartel model has become viable. In this model, users establish and maintain their social profiles and connections at a few of the dominant social sites and let those sites or third-party applications, which are developed specifically for those sites, fulfill their content needs. Facebook is the prime example of this model. The social sites in this model, like in all cartels, are the biggest winners here: they maintain full control over the social content graph and effectively determine which content users will have access to. Content sites in this model are reduced to social applications, with no ability to perform complex analysis on the underlying social graph. More importantly, they are also forced to adapt their user interaction experience to the overall user interaction theme of the host social sites. There are two major implications for users. First, users no longer have to maintain many social profiles and establish the same social connections at many different sites, which is a significant improvement over the Decentralized Model. Second, however, they are forced to have a central online social presence, without which they won't even have access to the contents otherwise would have been available on content sites.

|  | Factor | Decentralized Model | Closed Cartel | Open Cartel |
|---|---|---|---|---|
| Users | which site to interact with? | content site | social site | content site |
|  | multiple same connections and profiles? | yes | no | no |
| Content Sites | control over content | yes | limited | yes |
|  | control over social graph | yes | no | limited |
|  | control over activities | yes | no | yes |
| Social Sites | control over content | no | limited | no |
|  | control over social graph | no | yes | yes |
|  | control over activities | no | yes | limited |

**Table 2: A Comparison between Three Content Management Models for Social Content Sites.**

**Open Cartel Model**: The Open Cartel model is an integration between the Decentralized and Closed Cartel models. In this model, a few dominant social sites still maintain the social profiles and connections. However, through open standards, individual content sites are allowed to retrieve social information from them, given users' permission, and integrate it with the content they provide on their own sites. Furthermore, content sites are allowed to propagate social profiles and connections established on their own sites back to the social sites. Given this open access and depending on their levels of expertise, the content sites can now operate in one of the three levels. The simplest content sites can choose to delegate the management of both activities generated on their site and the social connections to the social sites. More sophisticated content sites can manage user activities on their own and simply rely on social sites to provide the social graph. Even more sophisticated content sites can maintain their own social graphs and keep them in sync with the social sites. These social graphs can be considered as focused views on the underlying global social graph. The implications for users are three-fold. First, similar to the Close Cartel model, there is no need for users to repeat their profiles and connections at many different places. Second, users will have multiple points of interaction where they can consume content powered by their social profiles and connections. Finally, user experience can be fully customized instead of conforming to the look and feel of the social sites.

**Discussion**: A summary of the comparison between the three models is listed in Table 2. While it is relatively clear that the Decentralized Model is being replaced by the Cartel models, word is still out on which Cartel model will eventually come out on top. The core issue here is control over the content and social activities. In the Closed Cartel model, content sites delegate the management of social activities and the presentation of content to the social sites and essentially become applications that can not survive without the host social site. In the Open Cartel model, social and content sites create symbiosis relationships, where social sites provide valuable information to enhance user experience on content sites, and content sites in turn realize the value of the social graphs on social sites and expand them by providing users with useful contents and engaging them in interesting activities. It is our belief that small niche content sites (e.g., your neighborhood reading group) will prefer the Closed Cartel model for ease of management, while larger content sites (e.g., New York Times and Y!Travel) would prefer the Open Cartel model.

## 6.2 Activity-Driven Data Management

A good understanding of social connections and activities can help cluster users and their associated contents in ways that would improve the data access performance. We briefly discuss how we can leverage those to cluster users for better query processing.

Consider a social content site similar to del.icio.us, where users connect with other users and tag items with tags. Let $\mathcal{U}$ be the set of user nodes. Given a $u \in \mathcal{U}$, we use $items(u)$ to denote items tagged by $u$, $network(u)$ to denote users connected to $u$, and $taggers(i,k)$ to denote users who tagged item $i$ with tag $k$.

**Queries and Scores:** For this study, we are interested in keyword-only queries, $Q_u = k_1, ..., k_n$. We first define the score of an item $i$ for user $u$ and a keyword $k_j$, $score_{k_j}(i,u) = f(network(u) \cap taggers(i,k_j))$, where $f$ is a monotone function. We further define the overall score of an item $i$ for a user query $Q_u$ as $score(i,u) = g(score_{k_1}(i,u), ..., score_{k_n}(i,u))$, where $g$ is a monotone aggregate function. While the framework is general enough to permit arbitrary monotone functions $f$ and $g$, we will use $f = count$ and $g = sum$, for ease of exposition.

**Indices:** Typically, in Information Retrieval, one inverted list index is created for each keyword [6]. Each entry in the list contains the identifier of a document along with its score for that keyword. Storing scores allows to sort entries in the inverted list thereby enabling top-$k$ pruning [16]. While in classic IR each document has a unique score for a keyword (e.g., tf*idf [6] or probabilistic [18]), in our problem, the score of an item for a tag depends on the network of the user *who* is asking the query.

One straightforward adaptation to our framework is to store one inverted list per *(tag, user)* pair and sort items in each list according to their scores for the tag and user. We denote such an index by $IL_k^u$, which contains entries of the form $(i, score_k(i,u))$. Each item will be replicated along with its score in each *(tag,user)* inverted list. At query time, items scores can be aggregated across all inverted lists relevant to query keywords. However, consider a moderately sized [19] social content site with $100,000$ users, 1 million items, and 1000 distinct tags. If on average each item receives 20 tags which are given by 5% of the users, the size of the index would be approximately 1 terabyte, assuming 10 bytes per index entry. This kind of space requirement can easily become prohibitive as the network and tagging activity expand.

**Clustering:** In [5], we explored user clustering strategies which achieve different compromises between storage space and processing time. Here, we formalize these strategies and expand them further. The intuitive idea is to cluster users according their social connections and activities such that score estimations can be done accurately without blowing up the index size. There are three main strategies: *network-based*, *behavior-based* and *hybrid*.

Given a cluster $C$, the score of an item $i$ in an index $IL_k^C$, is computed as the upper-bound of scores of $i$ for each user $u \in C$:

$$score_k(i, C) = max_{u \in C} score_k(i, u) \quad (1)$$

By storing score upper-bounds, top-$k$ pruning algorithms can still be used. However, score upper-bounds entail having to compute exact scores at query time for a specific user. This computation introduces some processing overhead compared with the straightforward approach, where exact scores are stored for each *(tag, user)* pair. To better understand this, we formalize the different user clustering methods.

DEFINITION 11 (NETWORK-BASED CLUSTER). *Two users $u_1$ and $u_2$ belong to the same network-based cluster if and only if the following predicate is true:*

$$\frac{|network(u_1) \cap network(u_2)|}{|network(u_1) \cup network(u_2)|} \geq \theta \quad (2)$$

where $\theta$ is an application-defined threshold. Two users fall into the same network-based cluster if their networks are similar enough. Given that item scores depend on user networks, it is natural to assume that an item would have a similar score for two users whose networks overlap substantially. Each user falls into a single cluster and an inverted list is created for each cluster, instead of each user.

In [5], we explored the space/time compromise of network-based clustering and showed that it consumes less space than the basic strategy without incurring too much query processing overhead. The applicability of network-based clustering to larger networks, obtained by integrating different social graphs, is the subject of future research.

Unfortunately, network-based clustering may have poor performance in the following scenario. Assume a user $u_1$ whose network contains users $v_1, v_2, ..., v_{20}$ and $v_{21}, ..., v_{25}$. Assume another user $u_2$ whose network contains $v_1, v_2, ..., v_{20}$ and that $u_1$ and $u_2$ end up in the same cluster. However, if most of the tagging actions come from users in $v_{21}, ..., v_{25}$, item scores for $u_1$ and $u_2$ would be very different. Clustering $u_1$ and $u_2$ would not be beneficial and would in fact incur unnecessary processing overhead. Consequently, we further explored behavior-based clustering.

DEFINITION 12 (BEHAVIOR-BASED CLUSTER). *Two users $u_1$ and $u_2$ belong to the same behavior-based cluster if and only if the following predicate is true:*

$$\frac{|items(u_1) \cap items(u_2)|}{|items(u_1) \cup items(u_2)|} \geq \theta \quad (3)$$

Here, two users belong to the same cluster if their tagging behavior is similar. In this case, the network members of a user $u$ may belong to multiple clusters. Therefore, at query time, potentially more clusters will be considered than in the network-based clustering strategy. In [5], we showed that behavior-based clustering achieves better processing time to the expense of space when compared to network-based clustering.

Ideally, one would want to combine the benefits of network-based and behavior-based clustering. We define hybrid clustering where two users fall into the same cluster if members of their network tag similarly. Here, we give the definition of a hybrid cluster: exploring the benefits of this strategy is the subject of future work.

DEFINITION 13 (HYBRID CLUSTER). *Two users $u_1$ and $u_2$ belong to the same hybrid cluster if and only if the following predicate is true:*

$$\frac{|items(v_1) \cap items(v_2)|}{|items(v_1) \cup items(v_2)|} \geq \theta \quad (4)$$

*for all users $v_1 \in network(u_1)$ and $v_2 \in network(u_2)$*

**Further Discussion**: We explored users' social connections and behaviors to answer a very simple kind of information discovery query: keyword-only queries. However, those social information can potentially be leveraged in many other fashions, including guiding information synchronization decisions from remote social sites. For example, a user who is highly connected may require more frequent synchronization of his network from social sites. The development of a framework to guide data storage and synchronization decisions based on users' social connections and activities is an interesting research field needs to be explored further.

## 7. INFORMATION PRESENTATION

Supporting effective user interactions in social content sites is not only a matter of locating relevant results for the user, but also identifying the right presentation of results. The right presentation can help a user explore the information more effectively, especially when she is not sure about exactly what she wants, which is often the case, as we learned from the `Y!Travel` queries. Our vision for the Information Presentation layer is to build *a dynamic result exploration framework*.

In search, presentation is primarily in the form of a single ranked list of results, where a result's rank reflects its degree of relevance to the input query. In recommender systems, presentation is an important aspect and has direct implications on building users' trust and giving them incentives to participate in more activities [24, 28]. There are many interesting new challenges in information presentation, including those that are related to user interface design. Here, we focus mainly on result grouping, and providing explanations for results and groups.

### 7.1 Grouping Items

Given a set of items $I_{Q_u}$ which have been computed for a user and a query, there are many different mechanisms for grouping items in $I_{Q_u}$: *Social Grouping*, which defines item groups based on similarity or closeness between users who endorsed the items; *Topical Grouping*, which defines item groups using the abstract topics each item belongs to; *Structural Grouping*, which relies on similarity in items' attributes. A key algorithmic challenge is *the dynamic discovery of groups* given a query result set $I_{Q_u}$. We provide here a formal definition of social grouping.

DEFINITION 14 (SOCIAL GROUPING). *Two items $i_1$ and $i_2$ belong to the same social group if and only if the following predicate is true:*

$$\frac{|taggers(i_1) \cap taggers(i_2)|}{|taggers(i_1) \cup taggers(i_2)|} \geq \theta \quad \Box \quad (5)$$

where $\theta$ is an application-specific threshold. The groups defined above are user-independent and could be pre-processed. When a query result $I_{Q_u}$ is computed, the task is to partition it into a set of *meaningful* groups. Group meaningfulness can be defined using a combination of the following criteria. First, *total number of groups*. Due to real estate on a page, the number of groups to display at a time needs to be restricted. Second, *group quality*, which is defined using the relevance of items in the group. Finally, *group size*, which is simply the number of items in the group.

Since screen real estate is limited, an interesting presentational alternative is to present the groups hierarchically, i.e., initially present a small number of groups appropriate for the screen area and upon request divide a group that the user is interested in into subgroups. Devising a grouping mechanism that dynamically adjusts with zoom-in and zoom-out requests is a promising presentation model that needs to further explored.

## 7.2 Explanations

Another challenge is to provide explanations on the results and descriptions of the groups. Unlike in web search, results from information discovery on social content sites are often endorsed by other users or are connected to other interesting objects, i.e., there exists a so-called *social provenance*. Letting users be aware of the social provenance often allows them to make more informed decisions as to what to do with the results. Similarly, providing descriptions on result groups can help them better understand the semantics behind those groups and therefore make better choices on what to explore further.

An explanation for a recommended item depends on the underlying recommendation strategy used [30]. If an item $i$ is recommended to user $u$ by a content-based strategy, then an *explanation* for recommendation $i$ is defined as:

$\text{Expl}(u, i) = \{i' \in \mathcal{I} \mid \text{ItemSim}(i, i') > 0 \;\&\; i' \in \text{Items}(u)\}$

i.e., the set of items similar to items ($i'$) that user $u$ has rated in the past. The explanation may contain more information such as the similarity weight $\text{ItemSim}(i, i') \times \text{rating}(u, i')$. Here, $\text{ItemSim}(i, i')$ returns a measure of similarity between two items $i$ and $i'$, and $\text{rating}(u, i')$ indicates the rating of item $i'$ by user $u$ (it is 0 if $u$ has not rated $i'$).

If an item $i$ is recommended to user $u$ by a collaborative filtering strategy, then an *explanation* for a recommendation $i$ is:

$\text{Expl}(u, i) = \{u' \in \mathcal{U} \mid \text{UserSim}(u, u') > 0 \;\&\; i \in \text{Items}(u')\}$

i.e., the set of users similar to $u$ who have rated item $i$. Similarly to item-based explanations, we can augment each user $u'$ in the explanation with the similarity weight $\text{UserSim}(u, u') \times \text{rating}(u', i)$. Here, $\text{UserSim}(u, u')$ returns a measure of similarity or connectivity between two users $u$ and $u'$ (it is 0 if $u$ and $u'$ are not connected).

In all cases, the explanation of a recommendation is either a set of items or a set of users, possibly together with weights as described above. Given an item explanation, there are many presentation alternatives. The most straightforward option is to list the set of users or items in the explanation of each item. Another alternative is to return aggregate information such as: "60% of your friends endorsed this item" or "This item is similar to 75% of items you visited before". The challenge is when and how to generated those aggregation information efficiently.

We can also define *group explanation*, $\text{Expl}(u, g)$, as an aggregation over individual item explanations in the group. However, it is more intriguing to explore how we can effectively convert individual explanations for items in a group into a concise explanation at a group level.

## 8. RELATED WORK

In a series of works, Mendelzon et al. [12, 11, 10] proposed query languages for manipulating graphs. The $G+$ language [12] was proposed as a complementary language for Datalog, for expressing recursive queries using visual concepts. Later, $G+$ was extended into $Hy+$ [11], a hypergraph-based visualization and querying language. In [10], additional primitives were added to support aggregation over edges as well as paths, without explicit recursion, but using transitive closure as a primitive. Amann and Scholl [4] proposed the Gram model and language for querying hypertext data modeled as graphs. The language includes limited support for recursion. All of these languages, however, use graph patterns extensively within their queries. This is in contrast to our algebraic approach, which relies on a set of operators that manipulate nodes and links.

In the context of semi-structured data, substantial work has been done on graph querying (e.g., Struql [17], UnQL [9], and Lorel [1]). Much of the emphasis was on querying graphs using regular path expressions over edge labels. Such expressions are too heavyweight for our applications. Finally, in the context of object-oriented databases, the GOOD data model and query language were developed by Paredaens et al. [20]. A key distinction between virtually every paper on graph querying and our work is that we do not expect the user to interact with the system using our query language. In addition, none of these previous works considers the integration of search, querying, and recommendation.

Indeed, while search and recommendation have been investigated separately, their combination has received very little attention, with perhaps the only exception in [15], where the authors studied the effectiveness of scoring functions in both search and recommendation. Another closely related work [26], which developed OLAP-style algorithms to answer social queries such as returning all the tags of a given user. Neither paper addresses the challenges of social content analysis, which is substantially more complex than queries.

Several approaches have been developed in the context of Web search result presentation. The approach in [25] is based on clustering results into groups of related topics. Gravano and Dakka [13] describe a hybrid method for summarizing online news articles which leverages pre-computation in order to efficiently compute document clusters, at query time. By contrast, our study focuses on result exploration through social, structural and topical groupings. In [22], the authors propose a presentation layer on top of a relational database in order to improve its usability, stressing the importance of provenance. While the idea of presentation is common to ours, their focus is not on information discovery over social content sites.

Finally, faceted search [14, 7] supports richer information discovery tasks over structured data. However, it mainly focuses on exposing hidden data correlations and providing aggregate counts along with each facet. It will be interesting to explore if social provenance can be considered with the faceted search framework.

## 9. CONCLUSION

We envision that domain-specific social content sites will increasingly become a part of the our online life. We motivated information discovery over such (real or virtual) social content sites and identified several major challenges. In particular, we proposed SocialScope, a logical architecture with three layers: Information Discovery, Content Management and Information Presentation. We discussed key issues and contributions in each layer.

In the context of Information Discovery, we proposed an algebraic framework to manipulate social content graphs. To the best of our knowledge, our algebra is the first one that is capable of manipulating social content graph in a uniform and flexible way. In the context of Content Management, we identified three main categories of data within social content sites: site content, social profiles and connections, and site-specific social activities. We examined three alternative content management models, each defined by how they management the three categories of data, and compared their benefits and drawbacks. We also discussed how to leverage common user behaviors to optimize data storage and indexing for query processing. Finally, in the context of Information Presentation, we discussed how novel ways of presenting information to users can help them understand the large variety of content discovered from social content sites.

We believe that SocialScope offers a framework in which key challenges in data management in social content sites can be addressed by our research community.